\documentclass[aps,floatfix,twocolumn,showpacs,showkeys,preprintnumbers,amsmath]{revtex4}
\usepackage{amssymb}
\usepackage[T1]{fontenc}
\usepackage[latin1]{inputenc}
\usepackage{graphicx}
\usepackage{dcolumn}
\usepackage{bm}

\begin{document}

\title{Superfluidity of $\Sigma^-$ hyperons in $\beta$-stable neutron star matter}

\author{Isaac Vida\~na}

\address{Gesellschaft f\"{u}r Schwerionenforschung (GSI), Planckstrasse 1, D-64291 Darmstadt, Germany}
 
\author{Laura Tol\'os}

\address{Institut f\"{u}r Theoretische Physik, J. W. Goethe-Universit\"{a}t, D-60054 Frankfurt am Main, Germany}

\date{\today}

\begin{abstract}

In this work we evaluate the $^1S_0$ energy gap of $\Sigma^-$ hyperons in $\beta$-stable neutron star matter. We solve the 
BCS gap equation for an effective $\Sigma^-\Sigma^-$ pairing interaction derived from the most recent parametrization of the 
hyperon-hyperon interaction constructed by the Nijmegen group. We find that the $\Sigma^-$ hyperons are in a $^1S_0$ superfluid
state in the density region $\sim 0.27-0.7$ fm$^{-3}$, with a maximum energy gap of order $8$ MeV at a total
baryon number density of $\sim 0.37$ fm$^{-3}$ and a $\Sigma^-$ fraction of about $8\%$. We examine the implications
on neutron star cooling.

\end{abstract}



\maketitle


Since the suggestion of Migdal \cite{mi60}, superfluidity in nuclear matter has received a great deal of attention 
over the last 40 years, partly due to its important consequences for a number of neutron star phenomena, such as  
pulsar glitches \cite{an75,sa89,pi92,sh83} and cooling rates \cite{sh83,pe92,pa94,el96b,sc98}. Nevertheless, whereas the presence of superfluid neutrons 
in the inner crust of neutron stars, and superfluid neutrons together with superconducting protons in their quantum 
fluid interior is well established and has been the subject of many studies \cite{ba90,ba92,ba92b,wa93,sc96,kh96,el96,lo01,de03,zu04}, 
a quantitative estimation of the pairing of other baryon species has not received so much attention up to date. In 
particular hyperons, which are expected to appear in neutron star matter at baryon number densities of order 
$\sim 2n_0$ ($n_0=0.17$ fm$^{-3}$), may also form superfluids if their interactions are attractive enough. 
It has been suggested that some neutron stars are cooled much faster than expected by a standard cooling mechanism 
(i.e., modified URCA processes), and that more rapid and efficient mechanisms are needed \cite{pa94,pr92,ts98,pa00,sc98b}. Processes
of the type $Y \rightarrow B+l+\bar{\nu}_l$ (e.g., $\Lambda \rightarrow p+e^-+\bar{\nu}_e ,\Sigma^- \rightarrow 
\Lambda +e^-+\bar{\nu}_e$, etc) can provide some of such rapid cooling mechanisms. Therefore, the study of hyperon superfluidity 
becomes of particular interest since it could play a key role in them. The case of $\Lambda$ superfluidity 
has been investigated by Balberg and Barnea \cite{bal98} using parametrized effective $\Lambda\Lambda$ interactions. 
Results for $\Lambda$ and $\Sigma^-$ pairing using several bare hyperon-hyperon interaction models have been recently presented 
by Takatsuka {\it et al.} \cite{ta99,ta00,ta02}. The results of both groups indicate the presence of a $\Lambda$ superfluid for 
baryon number densities in the range $2-4n_0$. The latter authors suggest that both $\Lambda$ and $\Sigma^-$ become
superfluid as soon as they appear in neutron star matter and that the formation of a $\Sigma^-$ superfluid may be more 
likely than that of a $\Lambda$ superfluid. 

Since the hyperon fraction ($n_Y/n_b$) in neutron star matter is not large ($10\%-30\%$ at most depending on the model), the 
Fermi momenta of hyperons are rather low although they appear at high values of the total baryon number densities. 
Therefore, the pairing interaction responsible for hyperon superfluidity, if it exists, should be that due to the 
$^1S_0$ wave which is most attractive at low momenta. In this work we evaluate the $^1S_0$ gap energies of $\Sigma^-$ 
hyperons in $\beta$-stable neutron star matter by solving the well-known BCS gap equation for an effective pairing interaction 
derived from the most recent parametrization of the free baryon-baryon potentials for the complete baryon octet as 
defined by Stoks and \mbox{Rijken} \cite{st99}. We employ the model NSC97e of this parametrization, since this model, 
together with the model NSC97f, results in the best predictions for hypernuclear observables \cite{ri98}.

The crucial quantity in determining the onset of superfluidity is the energy gap function $\Delta_k$. The value of 
this function at the Fermi surface is proportional to the critical temperature of the superfluid, and by determining 
it we therefore map out the region of the density--temperature plane where the superfluid may exist. To 
evaluate it we follow the scheme developed by Baldo {\it et al.} \cite{ba90}. These authors introduced an 
effective pairing interaction ${\tilde V}_{k,k'}$ defined according to
\begin{equation}
{\tilde V}_{k,k'}=V_{k,k'}-\sum_{k''>k_M}V_{k,k''}\frac{1}{2E_{k''}}{\tilde V}_{k'',k'} \ ,
\label{eq:vtild}
\end{equation}
which sums up all two-particle excitations above a cutoff momentum $k_M>k_F$ ($k_M=2$ fm$^{-1}$ in this work). Previous applications
of this method to the neutron and proton pairing \cite{ba90,el96} have shown that it is stable with respect to variations of $k_M$,
as we have also confirmed. The quasiparticle energy $E_k$ is given by $\sqrt{(\varepsilon(k)-\mu)^2+\Delta_k^2}$, being $\varepsilon(k)$ the 
single-particle energy in the medium for the particle species in question, $\mu$ the corresponding chemical potential, 
and $V_{k,k'}$ the free baryon-baryon potential in momentum space, in our case the bare $\Sigma^-\Sigma^-$ interaction
of the NSC97e baryon-baryon potential. We note that the $\Sigma^-\Sigma^-$ channel is purely isospin 2 and therefore there is no coupling
to other hyperon-hyperon channels in Eq.\ (\ref{eq:vtild}). For the $^1S_0$ channel the gap function can be determined by solving
\begin{equation}
\Delta_k=-\sum_{k'\leq k_M}{\tilde V}_{k,k'}\frac{\Delta_{k'}}{2E_{k'}} \ .
\label{eq:gap}
\end{equation}

\begin{figure}[t]
\centering
\includegraphics[angle=-90,scale=0.30]{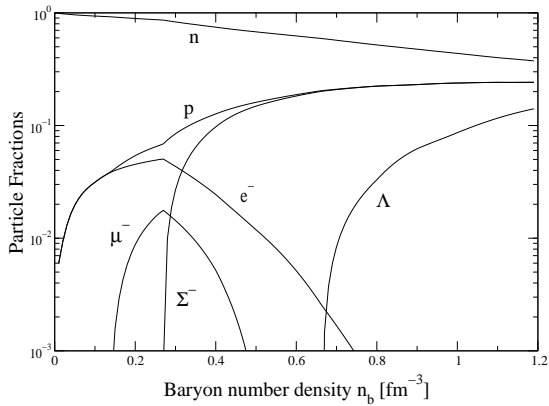}
\caption{Composition of $\beta$-stable neutron star matter. Taken from Ref.\ \cite{vi00}.}
\label{fig:fig1}
\end{figure}

Equations (\ref{eq:vtild}) and (\ref{eq:gap}) are solved self-consistently and represent a totally equivalent 
formulation of the BCS gap equation. With this procedure: (i) a well-behaved pairing interaction is obtained, since 
the repulsive core of the bare interaction is integrated out and (ii) double counting of two-particle correlations is
avoided. Excitations to intermediate states above $k_M$ are included in ${\tilde V}$, whereas excitations to states 
below $k_M$ are included in the gap equation (\ref{eq:gap}). We note here that for $k>k_F$ the dominant contribution 
to the quasiparticle energy $E_k$ comes from the term $(\varepsilon(k)-\mu)^2$. Therefore, we can neglect $\Delta_k$ 
in Eq.\ (\ref{eq:vtild}) for $k>k_M>k_F$. Thus Eq.\ (\ref{eq:vtild}) is decoupled from Eq. (\ref{eq:gap}), and we can solve the linear equation for 
${\tilde V}_{k,k'}$ by the matrix inversion method before proceeding to solve the gap equation by iteration (see Ref.\ \cite{el96} for 
details). 

The relevant $\Sigma^-$ fraction (shown in Fig.\ \ref{fig:fig1}), single-particle energy and chemical potential 
necessary to evaluate Eqs.\ (\ref{eq:vtild}) and (\ref{eq:gap}) are taken from the Brueckner--Hartree-Fock 
calculations described in Ref.\ \cite{vi00}, where the NSC97e baryon-baryon interaction was employed to describe the single-particle
properties, the composition and  Equation of State of $\beta$-stable neutron star matter, and the neutron star structure. Therefore, to our 
knowledge, the present work is the first one which employs consistently the same baryon-baryon interaction model 
to determine the single-particle properties, the composition, the Equation of State, the neutron star structure and the $\Sigma^-$ energy gap.

\begin{figure}[t]
\centering
\includegraphics[angle=-90,scale=0.30]{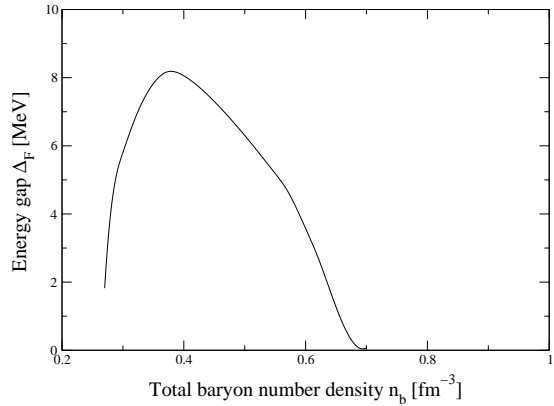}
\caption{Density dependence of the $\Sigma^-$ energy gap $\Delta_F$ in $\beta$-stable neutron star matter at $T=0$.}
\label{fig:fig2}
\end{figure}

Figure \ref{fig:fig2} shows the energy gap $\Delta_F$ of the $\Sigma^-$ hyperons in $\beta$-stable neutron star matter 
at $T=0$ with the composition shown 
in Fig.\ \ref{fig:fig1} as a function of the total baryon number density. Although, as can be seen in Fig.\ \ref{fig:fig1}, 
the $\Lambda$ may appear at higher densities, the $^1S_0$ $\Lambda\Lambda$ matrix elements of the Nijmegen interaction (NSC97a-f) are all 
weakly attractive, and therefore the energy gap for $\Lambda$ hyperons is expected to be zero at all densities, i.e., these 
particles will unlikely form a superfluid within our model. This is at variance with the results of Balberg and Barnea \cite{bal98}. Nevertheless, 
as stated before, these authors employed an effective parametrized interaction based on a $G$-matrix calculation to drive the gap equation and 
therefore overestimated, as pointed out by Takatsuka {\it et al.} \cite{ta99,ta00,ta02}, the $\Lambda$ energy gap mainly due to double counting effects. 
Our results for the $\Sigma^-$ are comparable to those of Takatsuka {\it et al.} \cite{ta00,ta02} which were obtained with several OBE 
hyperon-hyperon potentials. As these authors we find that $\Sigma^-$ hyperons are in a $^1S_0$ superfluid state as soon as they appear in matter and 
that the $\Sigma^-$ superfluid exists up to densities $\sim 4n_0$ with a critical 
temperature $T_c \sim 10^{10}$ K (see Fig.\ \ref{fig:fig4}). We find a maximum energy gap of order $8$ MeV at a total baryon number density of
$\sim 0.37$ fm$^{-3}$ and a $\Sigma^-$ fraction of about $8\%$. This gap is quite large in comparison with the neutron and proton ones
since the $\Sigma\Sigma$ (and in particular the $\Sigma^-\Sigma^-$) interaction in the Nijmegen NSC97a-f models is strongly attractive \cite{st99}. 
We want to emphasize, however, that this strong attraction is questionable.  Although these models
reproduce certain observables of $\Lambda$-hypernuclei, their predictions seem to be at odds with most of the scarce experimental data. The $\Lambda\Lambda$
interaction, as mentioned before, is weak compared to the values deduced experimentally \cite{ta01}, and all types of hyperons are too strongly 
bound in nuclear matter \cite{vi01}. This is especially suspect in the case of $\Sigma^-$, since phenomenology of $\Sigma^-$
atoms \cite{ma95} and hypernuclei \cite{da99} indicate a much weaker, if not repulsive, $\Sigma$ nuclear potential (see Ref.\ \cite{sch00} 
for a detailed discussion). Therefore, our results should be taken with caution.
 
\begin{figure}
\centering
\includegraphics[angle=-90,scale=0.30]{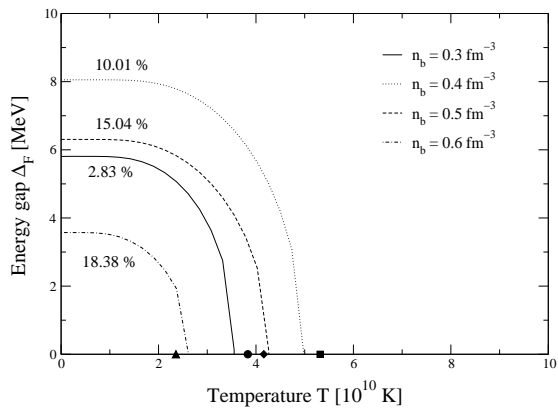}
\caption{Temperature dependence of the $\Sigma^-$ energy gap $\Delta_F$ in $\beta$-stable neutron star matter. The fraction of $\Sigma^-$ hyperon, $n_{\Sigma^-}/n_b$,
is indicated in each curve. The corresponding weak-coupling approximation
(WCA) estimations for the critical temperatures are also indicated by the circle ($n_b=0.3$ fm$^{-3}$), square ($n_b=0.4$ fm$^{-3}$), diamond ($n_b=0.5$ fm$^{-3}$) and
triangle ($n_b=0.6$ fm$^{-3}$).}
\label{fig:fig3}
\end{figure}

\begin{figure}
\vspace{0.5cm}
\centering
\includegraphics[angle=-90,scale=0.30]{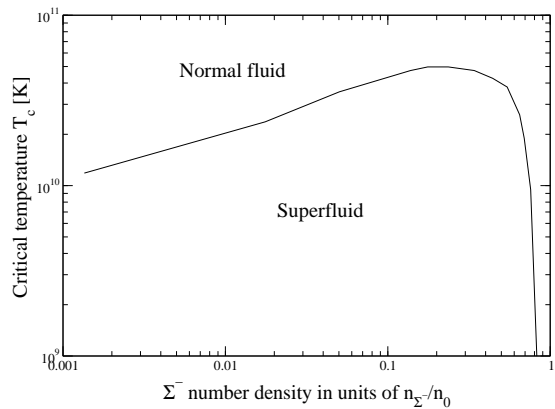}
\caption{Critical temperature of the $^1S_0$ $\Sigma^-$ superfluid as a function of the $\Sigma^-$ number density. The internal 
temperature of evolved normal neutron stars is around $10^8$ K. }
\label{fig:fig4}
\end{figure}

In Fig.\ \ref{fig:fig3} we show the temperature dependence of the energy gap $\Delta_F$ of $\Sigma^-$ for several values of
the total baryon number density and the corresponding $\beta$-stable fractions of the $\Sigma^-$. The gap function at finite temperature
can be obtained by solving
\begin{equation}
\Delta_k=-\sum_{k'\leq k_M}{\tilde V}_{k,k'}\frac{\Delta_{k'}}{2E_{k'}}\tanh\Big(\frac{E_{k'}}{2k_BT}\Big) \ ,
\label{eq:gap2}
\end{equation}
where $k_B$ is the Boltzmann's constant. We use the same approach as for the $T=0$ case. Here we ignore the temperature dependence
in ${\tilde V}_{k,k'}$ since for the temperature range of interest, $k_BT \approx 0-4$ MeV, the quasiparticle energy $E_K$ for $k>k_M$
is at least of order $100$ MeV, and thus we can ignore thermal excitations to states above $k_M$. In addition we use a ``frozen'' approximation
for the single-particle energy, chemical potential and fraction of the $\Sigma^-$, i.e., we use the corresponding quantities obtained
in the $T=0$ case, which is a reasonable approximation according to Refs.\ \cite{le86,ba88}. Therefore, as in the $T=0$ case, we first solve
Eq.\ (\ref{eq:vtild}) and then, with the effective interaction  ${\tilde V}_{k,k'}$ we solve Eq.\ (\ref{eq:gap2}). In Fig.\ \ref{fig:fig3}
we also show the critical temperatures estimated from the well-known weak-coupling approximation (WCA) \cite{li80}
\begin{equation}
k_BT_c\approx 0.57\Delta_F(T=0) \ ,
\label{eq:wca}
\end{equation}
which is a reasonable good approximation as can be seen from the figure.  

Finally, in Fig.\ \ref{fig:fig4} we show the region in the temperature--$\Sigma^-$-density plane where the $\Sigma^-$ hyperon is
expected to be superfluid. Since the values of the critical temperature are all well above the typical internal temperature of 
evolved normal neutron stars ($T_{int} \sim 10^8$ K), 
the $\Sigma^-$ is in a $^1S_0$ superfluid state for number densities ranging from $2.3 \times 10^{-4}$ fm$^{-3}$ up to $\sim 0.15$ fm$^{-3}$, 
which corresponds, according to the composition shown in Fig. \ref{fig:fig1}, to a total baryon number density ranging from
the $\Sigma^-$ onset density ($0.27$ fm$^{-3}$) to $\sim 0.7$ fm$^{-3}$ (see Fig. \ref{fig:fig2}).  

These results have implications for neutron star cooling. Since at low densities $\Sigma^-$ is the only hyperon species that is present 
in our model, the most important contribution to the neutrino cooling rate at such densities comes from the reaction 
$\Sigma^- \rightarrow n +e^-+\bar{\nu}_e$. In our model the threshold density for this reaction to occur is at around $1.6n_0$.
The direct action of such a rapid cooling mechanism, however, leads to surface temperatures much lower than that observed. Nevertheless,
if the $\Sigma^-$'s are in a superfluid state with energy gaps similar to what we found here, a sizeable reduction of the order $\exp(-\Delta_F/k_BT)$ may be 
expected in the neutrino emissivity of this process. Such a reduction will suppress the cooling rate and it will amount for neutron star surface 
temperatures more compatible with observation. Nevertheless, we should point out that this process will be also suppressed by the $^3P_2$ neutron pairing. This
pairing exists practically for all supernuclear densities \cite{am85} and, although it is relatively small ($\sim 0.1$ MeV), it will suppress this process throughout 
almost the entire life of the neutron star. We want to finish just mentioning that hyperon superfluidity may be also important for r-mode stability calculations, since
it may modify the temperature and density dependence of hyperon bulk viscosity \cite{li02}. 

\vspace{0.20cm}
The authors are very grateful to I. Bombaci, M. Hjorth-Jensen, A. Parre\~no, A. Polls, A. Ramos, J. Schaffner-Bielich 
and H.-J. Schulze for useful discussions, comments and critical reading of the manuscript. One of the authors (L.T.) wishes
to acknowledge the financial support from the Alexander von Humbolt Foundation.


\end{document}